\documentclass[runningheads,a4paper]{llncs}

\usepackage{amssymb}
\usepackage{graphicx}

\usepackage{color}
\usepackage{soul}
\usepackage{tikz}
\usepackage{amsmath}
\usepackage{amssymb}
\usepackage{mathtools}
\usepackage{pgfplots}
\usetikzlibrary{patterns}
\usetikzlibrary{calc}
\usetikzlibrary{pgfplots.groupplots}
\usetikzlibrary{backgrounds}
\usepackage[vlined,linesnumbered]{algorithm2e}

\usepackage{url}

\begin{document}

\mainmatter  

\title{LAMMPS' PPPM Long-Range Solver
  for the Second Generation Xeon Phi}


%
%
\author{William McDoniel\inst{1}\and Markus H\"ohnerbach\inst{1}\and Rodrigo Canales\inst{1}\\ Ahmed E. Ismail\inst{2}\and Paolo Bientinesi\inst{1}}
%
\authorrunning{W. McDoniel, M. H\"ohnerbach, R. Canales, A. E. Ismail, P. Bientinesi}

\institute{
RWTH Aachen University,
Aachen, Germany 52062\\
E-Mail: \texttt{mcdoniel@aices.rwth-aachen.de}
\and
West Virginia University,
Morgantown, USA 26506
}

%
%

\toctitle{Lecture Notes in Computer Science}
\tocauthor{Authors' Instructions}
\maketitle

\begin{abstract}
%

Molecular Dynamics is an important tool for computational biologists, chemists, and materials scientists, consuming a sizable amount of supercomputing resources.
Many of the investigated systems contain charged particles, which can only be simulated accurately using a long-range solver, such as PPPM.
We extend the popular LAMMPS molecular dynamics code with an implementation of PPPM particularly suitable for the second generation Intel Xeon Phi.
Our main target is the optimization of computational kernels by means of
vectorization, and we observe speedups in these kernels of up to 12x.
These improvements carry over to LAMMPS users, with overall speedups ranging between 2-3x, without requiring users to retune input parameters.
Furthermore, our optimizations make it easier for users to determine optimal input parameters for attaining top performance.

\end{abstract}

	\section{Introduction}\label{sec:introduction}
	Molecular dynamics simulations are used to compute the evolution of systems of atoms in fields as diverse as biology, chemistry, and materials science.
	Such simulations target millions or billions of particles, are frequently run in parallel, and consume a sizable portion of supercomputers' cycles. 
	Since in principle each atom interacts with all the other atoms in the system, 
	efficient methods to compute the pairwise forces are vital.
	The most widespread method for electrostatic interactions is the
	``Particle-Particle Particle-Mesh'' (PPPM) method~\cite{HockneyEastwood}, which makes
	it possible to efficiently compute even the interactions between distant particles.

	Due to its popularity, we target the open-source LAMMPS code~\cite{Sandia}, which offers the
	PPPM method.  LAMMPS is a C++ code designed for large parallel simulations
	using MPI, and is written to be modular and extensible.  LAMMPS can be
	compiled with a variety of packages that provide different implementations of key methods for the calculation of
	short-range and long-range interactions.  
	For example, the USER-OMP package includes versions
        of methods such as PPPM which are specifically designed for shared-memory parallelism.
	In this paper, we extend the LAMMPS molecular dynamics simulator
	with a version of PPPM that is especially suitable for architectures
        with wide vector registers, such as the 
	Xeon Phi.  In the past, long-ranged solvers have been optimized for GPUs, with
	issues similar to those encountered with Xeon Phi accelerators~\cite{BrownGPUPPPM,GPUSPME}.
	
        On these systems, one of the main routes towards high-performance is
        the exploitation of the wide (512-bit) vector registers.
	To this end, we  create vectorized kernels for all the computational
	components that are not directly supported by highly optimized math libraries (e.g. FFTs). 
	These  routines account for between 20\% and 80\% of the time spent in PPPM.
	As such, their optimization leads to notable speedups in the overall
        performance of the simulation.

	One challenge is that 
	the innermost loops of said computational routines are very short, with trip-counts between 3 and 7.  This is a common problem for vectorizing molecular dynamics even outside of PPPM.  For example, it was encountered by H\"ohnerbach et al. in their multi-platform vectorization of the extremely short loops of the Tersoff potential~\cite{MarkusSC16}. 
	It turns out that work can be saved elsewhere by increasing these trip counts, simultaneously enabling efficient vectorization.  Similarly, work can be shifted away from poorly-scaling FFTs and into newly-optimized functions, and, within the optimized functions, memory bandwidth can be traded against additional computation.
	

	In this paper, in addition to discussing vectorization techniques,  we 
        also provide insights into the parametrization of PPPM for performance.
	In particular, we consider three tunable
	parameters: the real-space cutoff, the interpolation order, and the differentiation mode.
	Many users will stick to the default choices where such exist, since these promise accurate and reasonably performant calculations.  Others will have taken time to tune these parameters for their particular problems, but even expert users often make suboptimal choices that can up to double time-to-solution for a given desired accuracy, depending on the problem~\cite{Diego16}.  We achieve 2-3x speedups for a wide range of input parameters, and our optimizations also make the careful tuning of several parameters unnecessary by making particular options superior to the others for almost all cases.
	
	The code presented in this paper is contributed to the USER-INTEL package of LAMMPS~\cite{BrownIntelLAMMPS}.
	It has been shown that this package can not just yield impressive speedups on Intel architecture, but also improve the energy efficiency of the calculation~\cite{BrownSC}.

	

	\section{Molecular Dynamics and PPPM}
        \subsection{An algorithmic overview}
	The interaction between atoms in a molecular dynamics simulation is governed by a so-called potential function.
	For example, the Lennard-Jones ($LJ$) and the
	Coulombic ($Coul$) potentials are given by:	
        \begin{equation}
          V_{LJ}^{ij} = 
          4\epsilon_{ij}\left[\left(\frac{\sigma_{ij}}{r_{ij}}\right)^{\mathrlap{12}}-\left(\frac{\sigma_{ij}}{r_{ij}}\right)^{6}\right],
          \quad \text{and} \quad
          V_{Coul}^{ij} = 
          \frac{C\ q_i\ q_j}{\varepsilon\ r_{ij}}.
          \label{eqn:coul}
        \end{equation}
	For a given pair of atoms $(i, j)$, the potential depends on the distance between them,
	$r_{ij}$, as well as their charges $q_i$ and $q_j$ (in the
        Coulombic case), 
        or the parameters $\epsilon_{ij}$ and $\sigma_{ij}$ (in the
        Lennard-Jones case), 
        which describe the minimum of the potential function and its root.
	In order to obtain the forces on atoms, MD simulations can compute
        these potential functions for all pairs of atoms, but
        $\mathcal{O}(n^2)$ pairs have to be evaluated, and this quickly becomes infeasible.
	

	A simple solution is to introduce a cutoff. One only considers interactions among atoms within a given cutoff radius $r_C$ of each other. Consequently, the number of pairs to be evaluated decreases to $\mathcal{O}(nr_C^3)$.
	Since all the potential functions (e.g., Eqn.~\ref{eqn:coul}) fall off with distance, the cutoff provides a reasonable
	strategy to approximate the total potential on atoms.
	
	There are, however, numerous situations in which long-range interactions between atoms
	cannot be neglected, and instead have to be approximated numerically. 
	A plain cutoff strategy does not work well for Coulomb interactions,
        which are relevant when a system contains charged particles or polar molecules,
        because the potential falls off only as $r^{-1}$.
	In contrast, the cutoff is perfectly fine for the Lennard-Jones potential, as long as the system is uniform.

	In non-uniform problems, such as those featuring an interface, even Lennard-Jones interactions may need to be calculated using a long-ranged solver and can not be approximated \cite{Veld}.
	In these cases, it is necessary to approximate these long-range interactions
	without explicitly computing pair-wise potential functions; for this task, 
	Particle-Particle Particle-Mesh is
	often the method of choice.
	%
	%
	%
	%
	%
	%
	PPPM approximates long-range interactions in a periodic system by obtaining the potential of the entire system of atoms as a function of space, discretized to a grid \cite{HockneyEastwood}.
	While originally developed for electrostatics, the method was later adapted
	to the $r^{-6}$ term of the Lennard-Jones potential \cite{IseleHolder}.
	
	In this work, we focus on PPPM for electrostatics, i.e., the Coulomb potential.
	PPPM uses an idea due to Ewald, and splits the potential into two components \cite{Ewald}.
	The first component, the ``short-ranged'' part of PPPM, contains  the discontinuity due to the $r^{-1}$ term, and
	a smooth screening term that limits the support to a small spherical region
	around a given atom;
	this component can be calculated directly between each atom and its
	neighbors in a certain cutoff radius $r_C$.
	The second component is the ``long-ranged'' part of PPPM; due to
	its smooth nature, this can be solved accurately on a grid.
	
	
	
	The efficient solution of the long-ranged component is
	the key ingredient 
	of the PPPM method.
	Since we are operating with smooth quantities, 
	the electrostatic potential is related to the charge distribution $\rho$ via Poisson's equation
	\begin{equation}\label{eqn:poisson}
	\nabla ^ 2 \Phi = - {\rho \over {\epsilon _0}}.
	\end{equation}	
	From the electrical potential $\Phi$, one can compute the forces on all the atoms due to it.
	The forces on an atom $j$ with charge $q_j$ can be obtained from the gradient of the potential evaluated at the particle's position:
	\begin{equation}\label{eqn:fpot}
	\vec{F}_j = - q_j \nabla \Phi.
	\end{equation}	
	PPPM approximates these forces on each particle by proceeding in three steps:
\begin{enumerate}
\item First, particle charges are mapped to a grid using a stencil,
  obtaining a discretized form of the charge distribution $\rho$. \label{itm:steps:1}
\item Second, Poisson's equation (Eqn.~\ref{eqn:poisson}) is \label{itm:steps:2}
  solved in order to obtain the potential $\Phi$.
  This is done by first taking the 3D Fourier transform of the charge
  distribution, as Poisson's equation is easier to solve in reciprocal
  space, and then performing one or more inverse FFTs to obtain a result in real space.
\item 
  Third, this result is mapped back to the atoms with the same stencil used when mapping charges. \label{itm:steps:3}
\end{enumerate}
	
The forces are obtained from the gradient of the potential, and this gradient can be taken in reciprocal or real space, determined by the user-specified differentiation mode.  For {\sc ik}  differentiation, the gradient is calculated in reciprocal space, immediately after solving Poisson's equation, and three inverse FFTs bring it back into real space, where its components are mapped to the atoms.  For {\sc ad} differentiation, one inverse FFT yields the scalar potential in real space, and this is mapped to the atoms using different sets of coefficients for each component of the gradient to be obtained.

	Our optimizations focus especially on the mapping steps (steps~\ref{itm:steps:1} and~\ref{itm:steps:3}).
	Step~\ref{itm:steps:2} is not as interesting for manual optimization since it is dominated by FFT calculations, for which highly optimized libraries exist.
	The mapping steps, on the contrary, are deeply nested loops performing calculations on data that is likely already in cache.
	We will show that optimizations, especially proper vectorization, will speed up these steps by at least a factor of four.

	
	
	
	\subsection{Related Work}
	
	Besides LAMMPS, many other popular molecular dynamics codes contain long-ranged solvers.
	Examples include, but are not limited to, Gromacs~\cite{Gromacs}, DL\_POLY~\cite{DLPOLY},
	AMBER~\cite{Amber}, Desmond~\cite{Desmond}, and NAMD~\cite{NAMD}.
	These codes tend not to implement PPPM itself, in favor of related
	schemes such as PME~\cite{PME}, SPME~\cite{SPME}, and
	$k$-GSE~\cite{GSE}.
	The main differences with respect to PPPM lie in the function used to
	interpolate atom charges onto the grid and back, and in the corresponding Green's function used to solve for the smooth part of the potential.
	There also exist schemes for long-ranged force evaluation that are not
	based on Fourier transforms, such as lattice Gaussian
	multigrid~\cite{LGM}, Multilevel Summation~\cite{Hardy}, and $r$-GSE~\cite{GSE}.
	
	\subsection{Parametrization of PPPM}\label{sec:profiling-lammps}
	
	Since LAMMPS is used for a wide variety of problems, users have many
	choices about input parameters for the target
	physical system.
	Several of these parameters influence the accuracy and/or speed of the
        simulation, including
	the cutoff distance 
        ($r_C$), the prescribed error in forces relative to a reference ($\epsilon$), the stencil
        size ($S$), and the differentiation mode, {\sc ik}  or {\sc ad}.  %
	
        $r_C$ expresses the distance
	within which pair-wise interactions are computed directly, and
	outside of which the interactions are approximated using the PPPM
	grid; the short-ranged calculations scale with $r_C^3$.  The work done when computing FFTs is controlled by $\epsilon$; LAMMPS automatically determines the coarseness of the FFT grid to satisfy this accuracy constraint, depending on the values chosen for the other parameters.
	A $7^3$ stencil ($S=7$) causes writing to, and reading from,
	about 2.7 times as many grid cells compared to the default $5^3$ stencil.  A
	higher-order stencil produces more accurate results, and LAMMPS takes this
	into account when deciding the resolution of the PPPM grid. Therefore, a
	higher-order stencil shifts work out of the FFT functions, and into the
	mapping functions.  Users can also choose between the {\sc ik}  and {\sc ad}
	differentiation modes described above, and LAMMPS again takes their
	different accuracies into account when setting up the FFT grid, with the  {\sc ik} mode yielding a slightly coarser grid.

	Users will typically want to use a set of inputs that nearly minimize
	runtime, subject to an accuracy constraint.  Unfortunately, short of
	trial-and-error for a specific problem it can be difficult to find a
	good set of parameters.  In a recent work~\cite{Diego16},
        Fabregat et al.~developed a method for
	automatically searching the space of input parameters to find a good set, guided
	by cost and accuracy models; 
	their case studies 
	suggest that even expert users systematically
	underestimate the expense of PPPM: they invariably predicted
	lower-than-optimal cutoffs, which minimize the work done in computing
	pair interactions while forcing a finer FFT grid. 
        The impact of stencil size was not considered, leaving the choice at LAMMPS'
	default.
	In the next sections we demonstrate that an appropriate choice of stencil size is needed to achieve good vectorization. 

	\subsection{Profiling}\label{sec:profiling} 
In order to investigate the effects of the input parameters on runtime, we execute our baseline on a single core of a KNL machine with a single thread.
The system is an Intel Xeon Phi 7210 chip (64 cores and 16GB of HBM RAM) in quadrant and flat memory mode, connected to other nodes via OmniPath.
Our software is based on the May 11, 2016 version of LAMMPS with the RIGID, USER-OMP and USER-INTEL packages enabled.
It was compiled using the Intel C++ Compiler version 16.01 (build 20151021), and uses Intel MPI 5.0 (build 20150128).
The reference runs use the code provided by the USER-OMP package, and our runs are based on code from USER-INTEL package running in mixed precision mode.
Our benchmark is an SPC/E water simulation~\cite{spce}, a benchmark provided with LAMMPS.
We modified it to have a cubic domain.

Since all the atoms in the system carry partial charges, the simulation uses PPPM to calculate forces.
Unless otherwise specified, the default settings that we use are relative error $\epsilon=10^{-4}$, and short-range cutoff $r_C=5$\AA.
The basecase contains 36,000 atoms, and will later be scaled up for more extensive benchmarks.

%

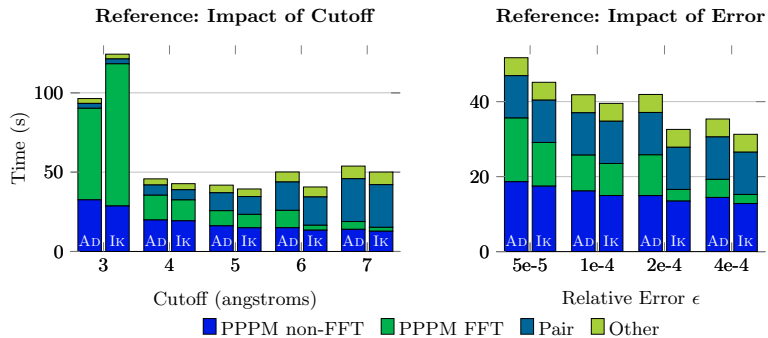
\begin{figure}[t]
\begin{center}
			\begin{tikzpicture}[scale=0.8]
			\begin{axis}[
			ybar stacked, bar width=11pt, bar shift = -6.5pt,
			xmin=-.5, xmax=4.5,
			ymin=0, ymax=130,
			width=7cm,
			height=5cm,
			xtick={0,1,2,3,4},
			xticklabels={3, 4, 5, 6, 7},
			legend style={at={(1.1,-0.30)},anchor=north,/tikz/every even column/.append style={column sep=.2cm}, anchor=north,draw=none, legend columns=4},
			ymajorgrids,
			ylabel={Time (s)},
			ylabel style={yshift=-.4cm},
			axis x line*=bottom,
			y axis line style={opacity=0},
			xlabel ={Cutoff (angstroms)},
			title={\textbf{Reference: Impact of Cutoff}}
			]
			\addplot[color=black, fill=blue!90!green] table[x expr=\coordindex, y index=0, header=false]{img/pppm_omp1_ik_cut.txt};
			\addplot[color=black, fill=blue!30!green] table[x expr=\coordindex, y index=1, header=false]{img/pppm_omp1_ik_cut.txt};		
			\addplot[color=black, fill=blue!60!green] table[x expr=\coordindex, y index=2, header=false]{img/pppm_omp1_ik_cut.txt};
			\addplot[color=black, fill=yellow!60!green] table[x expr=\coordindex, y index=3, header=false]{img/pppm_omp1_ik_cut.txt};
			\legend{PPPM non-FFT, PPPM FFT, Pair, Other};	
			\end{axis}
			
			\begin{axis}[
			ybar stacked, bar width=11pt, bar shift = 6.5pt,
			ymin=0, ymax=130,
			xmin=-.5, xmax=4.5,
			width=7cm,
			height=5cm,
			xtick={0,1,2,3,4},
			xticklabels={3, 4, 5, 6, 7},
			y axis line style={opacity=0},				
			]				
\pgfplotsinvokeforeach{0,1,2,3,4}{
	\node[anchor=south, white, scale=0.8] (A) at ($ (axis cs:#1,0.0) + 1*(6.5pt, 0)$) {\sc Ik};
	\node[anchor=south, white, scale=0.8] (A) at ($ (axis cs:#1,0.0) - 1*(6.5pt, 0)$) {\sc Ad};
}
			\addplot[color=black, fill=blue!90!green] table[x expr=\coordindex, y index=0, header=false]{img/pppm_omp1_ad_cut.txt};
			\addplot[color=black, fill=blue!30!green] table[x expr=\coordindex, y index=1, header=false]{img/pppm_omp1_ad_cut.txt};		
			\addplot[color=black, fill=blue!60!green] table[x expr=\coordindex, y index=2, header=false]{img/pppm_omp1_ad_cut.txt};
			\addplot[color=black, fill=yellow!60!green] table[x expr=\coordindex, y index=3, header=false]{img/pppm_omp1_ad_cut.txt};
			\end{axis}
\begin{scope}[xshift=7cm]
			\begin{axis}[
			ybar stacked, bar width=11pt, bar shift = -6.5pt,
			xmin=-.5, xmax=3.5,
			ymin=0, ymax=55,
			width=6cm,
			height=5cm,
			xtick={0,1,2,3},
			xticklabels={5e-5, 1e-4, 2e-4, 4e-4},
			legend style={at={(0.5,-0.2)},anchor=north,/tikz/every even column/.append style={column sep=0.2cm}, anchor=north,draw=none, legend columns=4},
			ymajorgrids,
			axis x line*=bottom,
			y axis line style={opacity=0},
			xlabel ={Relative Error $\epsilon$},
			title={\textbf{Reference: Impact of Error}}
			]
			\addplot[color=black, fill=blue!90!green] table[x expr=\coordindex, y index=0, header=false]{img/pppm_omp1_ik_acc.txt};
			\addplot[color=black, fill=blue!30!green] table[x expr=\coordindex, y index=1, header=false]{img/pppm_omp1_ik_acc.txt};		
			\addplot[color=black, fill=blue!60!green] table[x expr=\coordindex, y index=2, header=false]{img/pppm_omp1_ik_acc.txt};
			\addplot[color=black, fill=yellow!60!green] table[x expr=\coordindex, y index=3, header=false]{img/pppm_omp1_ik_acc.txt};
			\end{axis}
			
			\begin{axis}[
			ybar stacked, bar width=11pt, bar shift = 6.5pt,
			ymin=0, ymax=55,
			xmin=-.5, xmax=3.5,
			width=6cm,
			height=5cm,
			xtick={0,1,2,3},
			xticklabels={5e-5, 1e-4, 2e-4, 4e-4},
			y axis line style={opacity=0},				
			]				
\pgfplotsinvokeforeach{0,1,2,3,4}{
	\node[anchor=south, white, scale=0.8] (A) at ($ (axis cs:#1,0.0) + 1*(6.5pt, 0)$) {\sc Ik};
	\node[anchor=south, white, scale=0.8] (A) at ($ (axis cs:#1,0.0) - 1*(6.5pt, 0)$) {\sc Ad};
}
			\addplot[color=black, fill=blue!90!green] table[x expr=\coordindex, y index=0, header=false]{img/pppm_omp1_ad_acc.txt};
			\addplot[color=black, fill=blue!30!green] table[x expr=\coordindex, y index=1, header=false]{img/pppm_omp1_ad_acc.txt};		
			\addplot[color=black, fill=blue!60!green] table[x expr=\coordindex, y index=2, header=false]{img/pppm_omp1_ad_acc.txt};
			\addplot[color=black, fill=yellow!60!green] table[x expr=\coordindex, y index=3, header=false]{img/pppm_omp1_ad_acc.txt};
			\end{axis}
	%
\end{scope}
			\end{tikzpicture}
\end{center}
\vspace{-0.7cm}
			\caption{Profile of SPC/E water test case running single-threaded on one core of a KNL. Left bar: {\sc ad} differentiation, right bar: {\sc ik} differentiation.}
			\label{fig-cutoffomp1}
		\end{figure}

		Fig.~\ref{fig-cutoffomp1} shows timings as the cutoff, relative error, and differentiation mode vary.
The vertical sections denote the time spent in FFTs (``PPPM FFT''), and in
PPPM aside from FFTs (``PPPM non-FFT''), the time spent in the pair-wise short-ranged interactions (``Pair''), and everything else (``Other'').
		

For cutoff, there actually is a minimum of the runtime, i.e., reducing the cutoff will not reduce runtime beyond a certain point where the long-ranged part gets less efficient: The $r_C=3$\AA{} case spends a disproportionate amount of time in PPPM.
The cutoff mostly impacts the ``Pair'' time---since it scales as $\mathcal{O}(r_C^3)$---and the ``PPPM FFT'' time---since it forces the grid to grow or shrink.

For $\epsilon$, there of course is no minimum---lower accuracy results in faster simulations---mostly due to less time spent in FFT calculations (i.e. smaller grids).
Outliers in FFT performance can be attributed to pathological cases (in terms of size) of the FFT library.

In both panels of Fig.~\ref{fig-cutoffomp1}, the ``Other'' and the ``PPPM
non-FFT'' sections are largely unaffected by changes in cutoff or relative error.
In both, {\sc ad} differentiation performs best (except for one outlier).
For cutoff-optimal cases, the majority of the runtime is spent on long-ranged calculation, suggesting that optimization in that area might be quite fruitful.
%
	
		
\section{Optimizations}
The optimizations for the different stages of the algorithm are discussed
here. In particular, we cover the functions that map atoms to grid points and
grid values to atoms, the Poisson solver, and the routines responsible for the
short-ranged contribution.  

		\subsection{Mapping Functions} \label{sec:mapping-routines}

All three mapping functions---\emph{Map-Charge} and both the {\sc ik} and the
{\sc ad} versions of \emph{Distribute-Force}---share the same structure: a
loop over all atoms, the calculation of stencil coefficients, and then a loop over stencil points.
		\emph{Map-Charge} multiplies the particle charge by the stencil coefficient and adds that value to a point on the grid.
		\emph{Distribute-Force} proceeds in a slightly different way depending on the differentiation mode.
		The {\sc ik} mode 
		 multiplies the grid values for each spatial dimension at each
                 grid point by the corresponding stencil coefficient, then
                 adds them to three totals, one for each dimension;
                 after the loop over stencil points, these components are multiplied by the atom's charge and a scaling factor to obtain force components.  
		The {\sc ad} mode 
		multiplies the scalar potential at a grid point by three different stencil coefficients to obtain a vector, which is added on the atom;
		after the loop over stencil points, substantially more calculation than is required for {\sc ik} differentiation transforms these totals into the components of the force vector.



		The stencil coefficients are the product of three polynomials of order equal to the stencil size, one for each dimension.
		The iteration over stencil points consists of a triple loop
                (one for each dimension of the stencil).
This represents the bulk (80\%+) of the work, and accounts for almost all the
memory accesses in the mapping functions.  \emph{Map-Charge} accesses only a
single value at each grid point, but does very little computation.  The {\sc
  ik} mode of \emph{Distribute-Force} uses three different values at each grid
point. 
The {\sc ad} mode uses only one value at each grid point, but performs more
floating point operations.	The arithmetic intensity of all these routines
is relatively low, and memory access patterns will determine the best approach to vectorization.

Since the number of grid points is typically comparable to or smaller than the
number of atoms, and $NS^3$ stencil points are touched when looping over $N$
atoms, there is a great deal of data reuse.  With so few calculations being performed on data which is almost always found in cache, managing vectorization overhead will prove to be vital.
		In general, we find that it is important to minimize the amount of data shuffling or masking required to prepare for vector operations; whenever possible, a full vector should be pulled from memory, operated on, and returned.
		
\begin{figure} 
\begin{center}
	\begin{tikzpicture}[scale=0.8]
	\begin{axis}[ybar=0.2em, xtick={0,1,2}, 
	xticklabels={\emph{Map-Charge}, \emph{Distribute-Force} \sc{ik}, \emph{Distribute-Force} \sc{ad}}, 
	ymin=0, ymax=13,
	xmin=.25, xmax=1.75,
	legend style={font=\tiny},
	every x tick label/.append style={font=\small},
	enlarge x limits=0.5,
	width=11cm,
	height=5cm,
	bar width=0.35cm,
	legend columns=2,
	legend pos=north west,
	area legend,
	title={Speedup relative to USER-OMP},
	xlabel ={Function},]
	\addplot[color=black, fill=blue!90!green] table[x expr=\coordindex, y index=0, header=false]{img/subroutine_timing.txt};
	\addplot[color=black, fill=blue!30!green] table[x expr=\coordindex, y index=1, header=false]{img/subroutine_timing.txt};
	\addplot[color=black, fill=blue!60!green] table[x expr=\coordindex, y index=2, header=false]{img/subroutine_timing.txt};
	\addplot[color=black, fill=blue!60!red] table[x expr=\coordindex, y index=3, header=false]{img/subroutine_timing.txt};		
	\addplot[color=black, fill=blue!30!red] table[x expr=\coordindex, y index=4, header=false]{img/subroutine_timing.txt};		
	\node[rotate=90, anchor=west] (A) at ($ (axis cs:0,1.0) - 2*(0.2em, 0) - 2*(0.35cm, 0)$) {reference};
	\node[rotate=90, anchor=west] (A) at ($ (axis cs:0,2.0) - 1*(0.2em, 0) - 1*(0.35cm, 0)$) {rethread};
	\node[rotate=90, anchor=west] (A) at ($ (axis cs:0,4.2) - 0*(0.2em, 0) - 0*(0.35cm, 0)$) {vector};
	\node[rotate=90, anchor=east, white] (A) at ($ (axis cs:0,11.4) + 1*(0.2em, 0) + 1*(0.35cm, 0)$) {simd8};
	\node[rotate=90, anchor=east, white] (A) at ($ (axis cs:0,12.1) + 2*(0.2em, 0) + 2*(0.35cm, 0)$) {precompute};
	
	\node[rotate=90, anchor=west] (A) at ($ (axis cs:1,1.0) - 2*(0.2em, 0) - 2*(0.35cm, 0)$) {reference};
	\node[rotate=90, anchor=west] (A) at ($ (axis cs:1,1.8) - 1*(0.2em, 0) - 1*(0.35cm, 0)$) {atom simd};
	\node[rotate=90, anchor=west] (A) at ($ (axis cs:1,3.8) - 0*(0.2em, 0) - 0*(0.35cm, 0)$) {inner simd};
	\node[rotate=90, anchor=west] (A) at ($ (axis cs:1,4.1) + 1*(0.2em, 0) + 1*(0.35cm, 0)$) {repacking};
	\node[rotate=90, anchor=west] (A) at ($ (axis cs:1,4.4) + 2*(0.2em, 0) + 2*(0.35cm, 0)$) {precompute};
	
	\node[rotate=90, anchor=west] (A) at ($ (axis cs:2,1.0) - 2*(0.2em, 0) - 2*(0.35cm, 0)$) {reference};
	\node[rotate=90, anchor=west] (A) at ($ (axis cs:2,2.7) - 1*(0.2em, 0) - 1*(0.35cm, 0)$) {vector};
	\node[rotate=90, anchor=west] (A) at ($ (axis cs:2,4.1) - 0*(0.2em, 0) - 0*(0.35cm, 0)$) {split atom};
	\end{axis}
	\end{tikzpicture}
\end{center}
\vspace{-0.7cm}
	\caption{Speedups for different implementations of each of the three mapping functions relative to the USER-OMP baseline version.  
Charge mapping timings were obtained from simulations using {\sc ik} differentiation.}
	\label{fig-implementations}   
\end{figure}
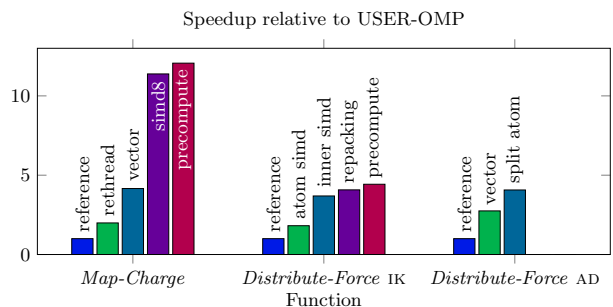

With an understanding of the structure of the mapping functions, we now walk through our process of optimizing each one, pointing out what worked and what did not.  A summary of progressive speedups for each function is shown in Fig. \ref{fig-implementations}.

\subsection*{Function \emph{Map-Charge}}
		

\hspace*{\parindent}{\bf Rethread:} To avoid race conditions when writing to the grid, the
USER-OMP package has threads own disjoint chunks of the grid, and uses
conditional statements in the innermost loop over stencil points. By giving
threads disjoint sets of atoms and maintaining private copies of the
grid---which are then summed together---we achieve a ${\sim}$2x speedup.
			
{\bf Vector:} We vectorize the innermost loop over stencil points, which features unit stride memory accesses as it iterates through grid points.  We target a new default stencil size of 7, instead of 5, to make better use of 256-bit vector registers.  This implementation achieves another factor of ${\sim}2$ speedup (``vector'' implementation), which is significant but not close to the theoretical 7x we might hope for.

{\bf Simd8:} Masking associated with the 7-iteration loop is a significant overheard.  By explicitly setting the loop length to 8 and padding the stencil coefficient arrays with zeros, we avoid masking and obtain a total of ${\sim}6$x speedup over the re-threaded scalar version.
			
{\bf Precompute:} Rather than evaluating polynomials to
                        obtain the stencil coefficients for each atom every
                        time step, we precompute 5000 values for each
                        polynomial and refer to the nearest entry in this
                        lookup table instead.  This brings total speedup to
                        over 12x of the baseline.

\subsection*{Function \emph{Distribute-Force} ({\sc ik} Differentiation)}

\hspace*{\parindent}{\bf Atom Simd:} 
Since \emph{Distribute-Force} performs reads from the grid rather than writes, the atom loop can be vectorized easily, yielding a ${\sim}2$x speedup.  The gather operations required to read grid point values cause this to be a poor choice.
	
{\bf Inner Simd:} Reproducing the inner loop vectorization from \emph{Map-Charge}, setting the loop length to 8, produces a ${\sim}3.7$x speedup over the scalar implementation.

{\bf Repacking:} \emph{Distribute-Force} for {\sc ik} differentiation uses three different grids with their own force components.  By modifying the Poisson solver to instead output the x and y components interweaved, and the z component interweaved with 0s, the innermost loop can be extended to 16 iterations and the x and y components can be computed together by taking advantage of the 512-bit vector register on Xeon Phi.  This provides an additional ${\sim}1.1$x speedup.

{\bf Precompute:} As with \emph{Map-Charge}, the polynomial evaluations to obtain stencil coefficients can be replaced with references to a lookup table, for a similar ${\sim}1.1$x additional speedup and a total speedup of ${\sim}4.4$x relative to the reference.

\subsection*{Function \emph{Distribute-Force}  ({\sc ad} Differentiation)}

\hspace*{\parindent}{\bf Vector:} Transferring over all of the optimizations from the {\sc ik} mode of \emph{Distribute-Force}, except the inapplicable repacking of the Poisson solver output, yields speedup below 3x relative to the reference.  This is because the extra work after the loop over stencil points has become relatively expensive.

{\bf Split Atom:} We split the loop over atoms in two.  The first atom loop ends after the triple loop over stencil points, having summed weighted potentials into three arrays of length equal to the number of atoms.
The second atom loop operates on these arrays to obtain force components, and can be vectorized as it contains no inner loops and has unit stride access to the weighted potential arrays.
This brings the overall speedup to just above 4x.

		\subsection{Poisson Solver}
The Poisson solver is a poorly-scaling, communication-intense function which performs 3D FFTs, solves Poisson's equation in reciprocal space, and then performs a number of inverse 3D FFTs depending on the differentiation mode (3 for {\sc{ik}} and 1 for {\sc{ad}}).  These 3D FFTs are performed in parallel as a series of 1D FFTs with communication steps in between.  The FFT functions are from high-performance libraries (in our case MKL) and we do not attempt to optimize them.
Our optimization of the solver comes from three ideas.

{\bf{Shift Work:}} Switching to a stencil size of 7 creates more work in the mapping functions, but causes LAMMPS to choose a coarser grid resolution, requiring fewer calculations to perform the FFTs.

{\bf{2D FFTs:}} The series of 1D FFTs is inefficient \cite{wendecug16}.  We replace it with a 2D FFT followed by a 1D FFT, and in the first communication step we ensure that planes of data are located on each MPI rank.  This saves one communication step and is roughly (${\sim}10\%$) faster.  Even for poorly load-balanced cases, where the number of necessary 2D FFTs is only slightly greater than the number of MPI ranks, it does not perform worse.

{\bf Adjust Grid Sizes:} The FFT calls of Intel's MKL library do not perform well for particular unfortunate values, which can catch users by surprise (compare time spent in FFTs across the cases in Fig. \ref{fig-cutoffomp1}). 
A simple fix that catches many problem cases is to check
whether the number FFT grid points in any dimension is a
multiple of 16, and  increase it by 1 if necessary.  Users will now only rarely find that their simulations run substantially slower after making a tiny change to their input file, and, as an added bonus, these simulations will gain slightly improved accuracy.

		\subsection{Short-Ranged Interactions}
		

		To avoid shifting the bottleneck to the short-range calculation, it is desirable that it be vectorized.
		Mike Brown of Intel contributed code vectorizing the pair potential used in simulations containing electrostatic interactions (optionally with cut off Lennard-Jones interactions), where his strategy was to vectorize the loop over each atom's neighbors.  
		This achieves a ${\sim}3$x speedup (for example, compare the time spent in ``Pair" between the reference and optimized versions in Fig. \ref{fig-1xstencil_compare}). 
		We provide similar code compatible with the Buckingham potential, optimized for PPPM and USER-INTEL, and also versions of pair potentials compatible with PPPM for dispersion.
		
		\section{Results}\label{sec:results}
		
		We now present comparisons between the reference and optimized
		versions of LAMMPS using full simulations, profiling the code as in
		Fig.~\ref{fig-cutoffomp1}, to show how the various parts of the code
		contribute to total runtime.  
		We also investigate the opaque way in which the user-facing knobs impact
		accuracy, and provide evidence that 
		our optimizations do not sacrifice accuracy. The experiments were conducted on a single core, a full node, and multiple nodes.
		While the speedup is both problem dependent and parameter
		dependent, the optimized version is faster in
		every case simulated.
		
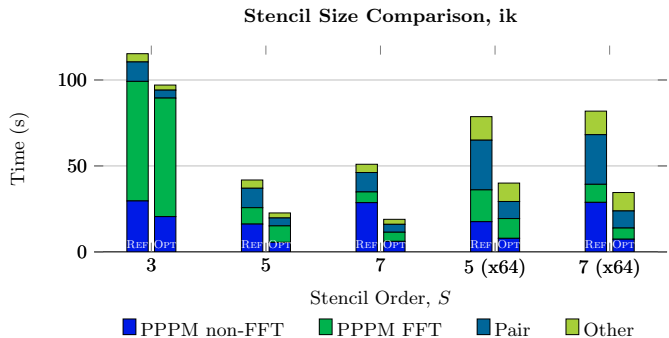
\begin{figure}
\centering
			\begin{tikzpicture}[scale=0.8]
			\begin{axis}[
			ybar stacked, bar width=10pt, bar shift = -6.5pt,
			xmin=-.5, xmax=4.5,
			ymin=0, ymax=120,
			width=11cm,
			height=5cm,
			xtick={0,1,2,3,4},
			xticklabels={3,5,7, 5 (x64), 7 (x64)},
			legend style={at={(0.5,-0.3)},anchor=north,/tikz/every even column/.append style={column sep=0.5cm}, anchor=north,draw=none, legend columns=4},
			ymajorgrids,
			ylabel={Time (s)},
			axis x line*=bottom,
			y axis line style={opacity=0},
			xlabel ={Stencil Order, $S$},
			title={\textbf{Stencil Size Comparison, ik}}
			]
			\addplot[color=black, fill=blue!90!green] table[x expr=\coordindex, y index=0, header=false]{img/pppm_omp_stencilvary.txt};
			\addplot[color=black, fill=blue!30!green] table[x expr=\coordindex, y index=1, header=false]{img/pppm_omp_stencilvary.txt};		
			\addplot[color=black, fill=blue!60!green] table[x expr=\coordindex, y index=2, header=false]{img/pppm_omp_stencilvary.txt};
			\addplot[color=black, fill=yellow!60!green] table[x expr=\coordindex, y index=3, header=false]{img/pppm_omp_stencilvary.txt};	
			\legend{PPPM non-FFT, PPPM FFT, Pair, Other}				
			\end{axis}
			\begin{axis}[
			ybar stacked, bar width=10pt, bar shift = 6.5pt,
			xmin=-.5, xmax=4.5,
			ymin=0, ymax=120,
			width=11cm,
			height=5cm,
			xtick={0,1,2,3,4},
			xticklabels={3,5,7, 5 (x64), 7 (x64)},
			y axis line style={opacity=0},				
			]				\pgfplotsinvokeforeach{0,1,2,3,4}{
	\node[anchor=south, white, scale=0.6] (A) at ($ (axis cs:#1,0.0) + 1*(6.5pt, 0)$) {\sc Opt};
	\node[anchor=south, white, scale=0.6] (A) at ($ (axis cs:#1,0.0) - 1*(6.5pt, 0)$) {\sc Ref};
}
			\addplot[color=black, fill=blue!90!green] table[x expr=\coordindex, y index=0, header=false]{img/pppm_new_stencilvary.txt};
			\addplot[color=black, fill=blue!30!green] table[x expr=\coordindex, y index=1, header=false]{img/pppm_new_stencilvary.txt};		
			\addplot[color=black, fill=blue!60!green] table[x expr=\coordindex, y index=2, header=false]{img/pppm_new_stencilvary.txt};
			\addplot[color=black, fill=yellow!60!green] table[x expr=\coordindex, y index=3, header=false]{img/pppm_new_stencilvary.txt};
			\end{axis}

			\end{tikzpicture}				
\vspace{-0.5cm}
\caption{Profiles of SPC/E water as stencil size varies for both single-core and scaled-up full-node cases.  Left bar: reference, right bar: optimized.}
			\label{fig-1xstencil_compare}
		\end{figure}

		Because of our decision to target a new default stencil size of 7, it would
		not be fair to make like-to-like comparisons between the reference and our
		optimized versions.  Further, LAMMPS' input files do not even require an
		explicit choice of stencil size, so many users will just allow it to take on
		its default value. 
%
		Fig.~\ref{fig-1xstencil_compare} compares the two versions as stencil order
		varies for our baseline test cases, using {\sc ik} differentiation to demonstrate that the new value is faster for the optimized version.  
		We simulate the standard 5\AA\ case on a single core and a 64x scaled-up 7\AA\
		case on a full KNL node, which are nearly-optimal cutoff radii for each case. 
%
		The trend in total runtime is expected: on both a single core and the full node the reference version is fastest with a stencil size of 5 while the new version is fastest with a value of 7.  For all future cases presented, the reference code uses $S=5$ while ours uses $S=7$.
		
		
\subsection{Accuracy}\label{sec:accuracy}
		
Since the optimizations proposed involve both parameter-tuning and numerical
approximations, we now verify that our code is as accurate as the
reference. To this end, we 
compare to an Ewald summation run with a relative error of
$10^{-5}$, and a cutoff of 10\AA.

\begin{table}[!t]
  \begin{center}
    \begin{tabular}{c c @{\ \ } l @{\ \ }  l @{\ \ } r @{\ \ } c  || c  c  @{\ \ } l @{\ \ } l @{\ \ } c }
      Version & mode & $r_c$ & $S$ & precompute & RMS error & Version & mode & $r_c$ & $S$ &  RMS error \\
      \hline
      ref & {\sc{ik}} & 7\AA & 7 & \hfill-\hfill & 0.0186 & ref & {\sc{ad}} & 7\AA & 7  & 0.0189	\\          
      opt & {\sc{ik}} & 7\AA & 7 & -             & 0.0186 & ref & {\sc{ik}} & 3\AA & 7 & 0.5853 \\      
      opt & {\sc{ik}} & 7\AA & 7 & 500 points    & 0.0313 & ref & {\sc{ik}} & 5\AA & 7 &  0.0124 \\ 
      opt & {\sc{ik}} & 7\AA & 7 & 5000 points   & 0.0188 & ref & {\sc{ik}} & 7\AA & 3 &  0.0197 \\
      opt & {\sc{ad}} & 7\AA & 7 & 5000 points   & 0.0188 & ref & {\sc{ik}} & 7\AA & 5  & 0.0194 \\  
    \end{tabular}
    \caption{RMS errors for force after one timestep compared to Ewald summation}
  \end{center}
  \label{accuracy_table}	
\end{table}

As seen in Table \ref{accuracy_table}, without
stencil coefficient precomputation, the optimized and reference versions
obtain almost identical forces for both differentiation modes.  5000
precomputed stencil polynomial evaluations are sufficient to retain overall
accuracy with our approximation.  In addition, the optimized version conserves
momentum (the sum of forces on all atoms remains nearly zero) and the
macroscopic temperature difference between reference and optimized simulations
after 100 time steps is always small (${\sim}0.1\%$), and nearly zero without stencil precomputation.

Many users may not expect that their choice of cutoff can have a large effect
on accuracy, and LAMMPS' internal accuracy model does not do as good of a job
with stencil size as it does with differentiation mode. After 100 time
steps, the temperature is almost 10 degrees higher for a 3\AA\ cutoff than for
cutoffs greater than or equal to 4\AA.  In addition to speedup, our optimized version becomes slightly more accurate by moving to a stencil size of 7.

\subsection{Single-Core Simulations}
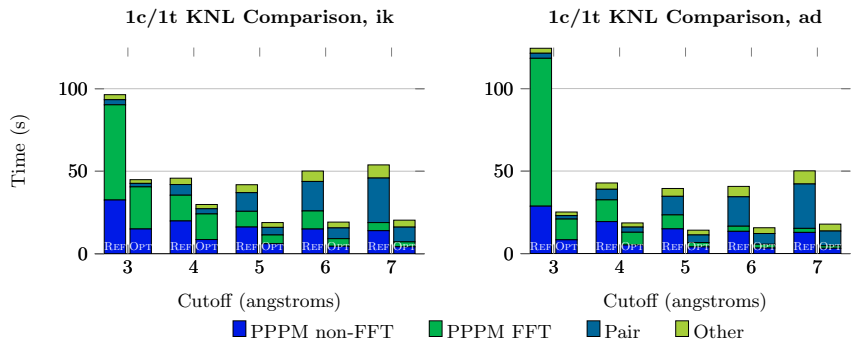
\begin{figure}
			\centering
			\begin{tikzpicture}[scale=0.8]
			\begin{axis}[
			ybar stacked, bar width=10pt, bar shift = -6pt,
			ymin=0, ymax=125,
			xmin=-.5, xmax=4.5,
			width=7cm,
			height=5cm,
			xtick={0,1,2,3,4},
			xticklabels={3, 4, 5, 6, 7},
			legend style={at={(1.2,-0.30)},anchor=north,/tikz/every even column/.append style={column sep=0.5cm}, anchor=north,draw=none, legend columns=4},
			ymajorgrids,
			ylabel={Time (s)},
			axis x line*=bottom,
			y axis line style={opacity=0},
			xlabel ={Cutoff (angstroms)},
			title={\textbf{1c/1t KNL Comparison, ik}}
			]
			\addplot[color=black, fill=blue!90!green] table[x expr=\coordindex, y index=0, header=false]{img/pppm_omp1_ik_cut.txt};
			\addplot[color=black, fill=blue!30!green] table[x expr=\coordindex, y index=1, header=false]{img/pppm_omp1_ik_cut.txt};		
			\addplot[color=black, fill=blue!60!green] table[x expr=\coordindex, y index=2, header=false]{img/pppm_omp1_ik_cut.txt};
			\addplot[color=black, fill=yellow!60!green] table[x expr=\coordindex, y index=3, header=false]{img/pppm_omp1_ik_cut.txt};
			\legend{PPPM non-FFT, PPPM FFT, Pair, Other}				
			\end{axis}
			
			\begin{axis}[
			ybar stacked, bar width=10pt, bar shift = 6pt,
			ymin=0, ymax=125,
			xmin=-.5, xmax=4.5,
			width=7cm,
			height=5cm,
			xtick={0,1,2,3,4},
			xticklabels={3, 4, 5, 6, 7},
			y axis line style={opacity=0},				
			]				
				\pgfplotsinvokeforeach{0,1,2,3,4}{
	\node[anchor=south, white, scale=0.6] (A) at ($ (axis cs:#1,0.0) + 1*(6pt, 0)$) {\sc Opt};
	\node[anchor=south, white, scale=0.6] (A) at ($ (axis cs:#1,0.0) - 1*(6pt, 0)$) {\sc Ref};
}

			\addplot[color=black, fill=blue!90!green] table[x expr=\coordindex, y index=0, header=false]{img/pppm_new1_ik_cut.txt};
			\addplot[color=black, fill=blue!30!green] table[x expr=\coordindex, y index=1, header=false]{img/pppm_new1_ik_cut.txt};	
			\addplot[color=black, fill=blue!60!green] table[x expr=\coordindex, y index=2, header=false]{img/pppm_new1_ik_cut.txt};
			\addplot[color=black, fill=yellow!60!green] table[x expr=\coordindex, y index=3, header=false]{img/pppm_new1_ik_cut.txt};
			\end{axis}
			\begin{scope}[xshift=7cm]
			\begin{axis}[
			ybar stacked, bar width=10pt, bar shift = -6pt,
			ymin=0, ymax=125,
			xmin=-.5, xmax=4.5,
			width=7cm,
			height=5cm,
			xtick={0,1,2,3,4},
			xticklabels={3, 4, 5, 6, 7},
			legend style={at={(0.5,-0.2)},anchor=north,/tikz/every even column/.append style={column sep=0.5cm}, anchor=north,draw=none, legend columns=2},
			ymajorgrids,
			axis x line*=bottom,
			y axis line style={opacity=0},
			xlabel ={Cutoff (angstroms)},
			title={\textbf{1c/1t KNL Comparison, ad}}
			]
			\addplot[color=black, fill=blue!90!green] table[x expr=\coordindex, y index=0, header=false]{img/pppm_omp1_ad_cut.txt};
			\addplot[color=black, fill=blue!30!green] table[x expr=\coordindex, y index=1, header=false]{img/pppm_omp1_ad_cut.txt};	
			\addplot[color=black, fill=blue!60!green] table[x expr=\coordindex, y index=2, header=false]{img/pppm_omp1_ad_cut.txt};
			\addplot[color=black, fill=yellow!60!green] table[x expr=\coordindex, y index=3, header=false]{img/pppm_omp1_ad_cut.txt};
			\end{axis}
			
			\begin{axis}[
			ybar stacked, bar width=10pt, bar shift = 6pt,
			ymin=0, ymax=125,
			xmin=-.5, xmax=4.5,
			width=7cm,
			height=5cm,
			xtick={0,1,2,3,4},
			xticklabels={3, 4, 5, 6, 7},
			y axis line style={opacity=0},				
			]	
				\pgfplotsinvokeforeach{0,1,2,3,4}{
	\node[anchor=south, white, scale=0.6] (A) at ($ (axis cs:#1,0.0) + 1*(6pt, 0)$) {\sc Opt};
	\node[anchor=south, white, scale=0.6] (A) at ($ (axis cs:#1,0.0) - 1*(6pt, 0)$) {\sc Ref};
}

			\addplot[color=black, fill=blue!90!green] table[x expr=\coordindex, y index=0, header=false]{img/pppm_new1_ad_cut.txt};
			\addplot[color=black, fill=blue!30!green] table[x expr=\coordindex, y index=1, header=false]{img/pppm_new1_ad_cut.txt};
			\addplot[color=black, fill=blue!60!green] table[x expr=\coordindex, y index=2, header=false]{img/pppm_new1_ad_cut.txt};
			\addplot[color=black, fill=yellow!60!green] table[x expr=\coordindex, y index=3, header=false]{img/pppm_new1_ad_cut.txt};
			\end{axis}
			\end{scope}
			\end{tikzpicture}	
			\vspace{-0.5cm}			
			\caption{Profiles of SPC/E water test case running single-threaded on one core of a KNL as cutoff varies. Left bar: reference, right bar: optimized.}
			\label{fig-1xadcut_compare}
		\end{figure}

We first compare simulations using our optimized version to the reference cases we
presented earlier in Fig. \ref{fig-cutoffomp1}. 
Fig.~\ref{fig-1xadcut_compare} shows both versions as cutoff varies for {\sc ik} and {\sc ad} differentiation, respectively.
As with the reference version, there is a runtime-optimal
cutoff for the optimized version at 5\AA\ where a balance is struck
between the pair interactions and the FFTs.  Total speedup at this
optimal cutoff is 2.21x for {\sc ik} and 2.75x for {\sc ad} differentiation.
With our optimizations, {\sc ad}
differentiation goes from being only marginally faster at the runtime-optimal cutoff to being 32\%
faster, making it a compelling choice even for serial simulations
where the FFTs do not take up much time.  

The calculation of long-range interactions, inclusive of the mapping
functions, the FFTs, and various minor functions (PPPM FFT plus PPPM non-FFT),
is sped up by a factor of 3.44x for {\sc ad} differentiation.  The calculation of
the long-range interactions \textit{excluding} the FFTs has actually sped up
by a higher factor of 3.61x despite the larger stencil requiring looping over
2.74 times as many grid points.  The calculation of pair interactions is sped
up by about 2.5x.  {\sc ad} differentiation is now faster than {\sc ik} differentiation
for every cutoff, due to the smooth decrease in time spent performing FFTs as
cutoff increases.

The relative penalty for choosing a poor cutoff has not changed much except for cases where an unfortunate number of FFT grid points was doubling the cost of FFTs.  In general, an overestimate of the runtime-optimal cutoff is much less penalizing than an underestimate because the cost of the FFTs increases rapidly as cutoff decreases.  Because the optimized long-range calculations are sped up by about as much as the optimized short-range calculations, users will find that pre-existing input files and intuitions about runtime-optimal cutoffs still yield good results.

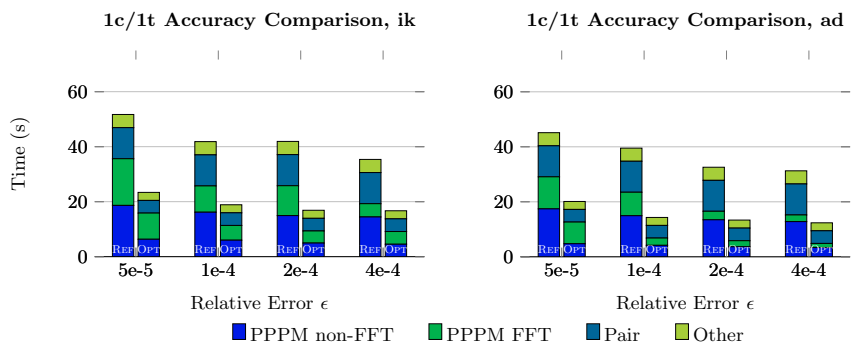
\begin{figure}
\centering
			\begin{tikzpicture}[scale=0.8]
			\begin{axis}[
			ybar stacked, bar width=10pt, bar shift = -6pt,
			xmin=-.5, xmax=3.5,
			ymin=0, ymax=75,
			width=7cm,
			height=5cm,
			xtick={0,1,2,3},
			xticklabels={5e-5, 1e-4, 2e-4, 4e-4},
			legend style={at={(1.2,-0.30)},anchor=north,/tikz/every even column/.append style={column sep=0.5cm}, anchor=north,draw=none, legend columns=4},
			ymajorgrids,
			ylabel={Time (s)},
			axis x line*=bottom,
			y axis line style={opacity=0},
			xlabel ={Relative Error $\epsilon$},
			title={\textbf{1c/1t Accuracy Comparison, ik}}
			]
			\addplot[color=black, fill=blue!90!green] table[x expr=\coordindex, y index=0, header=false]{img/pppm_omp1_ik_acc.txt};
			\addplot[color=black, fill=blue!30!green] table[x expr=\coordindex, y index=1, header=false]{img/pppm_omp1_ik_acc.txt};		
			\addplot[color=black, fill=blue!60!green] table[x expr=\coordindex, y index=2, header=false]{img/pppm_omp1_ik_acc.txt};
			\addplot[color=black, fill=yellow!60!green] table[x expr=\coordindex, y index=3, header=false]{img/pppm_omp1_ik_acc.txt};
			\legend{PPPM non-FFT, PPPM FFT, Pair, Other}				
			\end{axis}
			
			\begin{axis}[
			ybar stacked, bar width=10pt, bar shift = 6pt,
			ymin=0, ymax=75,
			xmin=-.5, xmax=3.5,
			width=7cm,
			height=5cm,
			xtick={0,1,2,3},
			xticklabels={5e-5, 1e-4, 2e-4, 4e-4},
			y axis line style={opacity=0},				
			]				
	
				\pgfplotsinvokeforeach{0,1,2,3,4}{
	\node[anchor=south, white, scale=0.6] (A) at ($ (axis cs:#1,0.0) + 1*(6pt, 0)$) {\sc Opt};
	\node[anchor=south, white, scale=0.6] (A) at ($ (axis cs:#1,0.0) - 1*(6pt, 0)$) {\sc Ref};
}

			\addplot[color=black, fill=blue!90!green] table[x expr=\coordindex, y index=0, header=false]{img/pppm_new1_ik_acc.txt};
			\addplot[color=black, fill=blue!30!green] table[x expr=\coordindex, y index=1, header=false]{img/pppm_new1_ik_acc.txt};		
			\addplot[color=black, fill=blue!60!green] table[x expr=\coordindex, y index=2, header=false]{img/pppm_new1_ik_acc.txt};
			\addplot[color=black, fill=yellow!60!green] table[x expr=\coordindex, y index=3, header=false]{img/pppm_new1_ik_acc.txt};
			\end{axis}
\begin{scope}[xshift=7cm]
			\begin{axis}[
			ybar stacked, bar width=10pt, bar shift = -6pt,
			xmin=-.5, xmax=3.5,
			ymin=0, ymax=75,
			width=7cm,
			height=5cm,
			xtick={0,1,2,3},
			xticklabels={5e-5, 1e-4, 2e-4, 4e-4},
			legend style={at={(0.5,-0.2)},anchor=north,/tikz/every even column/.append style={column sep=0.5cm}, anchor=north,draw=none, legend columns=2},
			ymajorgrids,
			axis x line*=bottom,
			y axis line style={opacity=0},
			xlabel ={Relative Error $\epsilon$},
			title={\textbf{1c/1t Accuracy Comparison, ad}}
			]
			\addplot[color=black, fill=blue!90!green] table[x expr=\coordindex, y index=0, header=false]{img/pppm_omp1_ad_acc.txt};
			\addplot[color=black, fill=blue!30!green] table[x expr=\coordindex, y index=1, header=false]{img/pppm_omp1_ad_acc.txt};		
			\addplot[color=black, fill=blue!60!green] table[x expr=\coordindex, y index=2, header=false]{img/pppm_omp1_ad_acc.txt};
			\addplot[color=black, fill=yellow!60!green] table[x expr=\coordindex, y index=3, header=false]{img/pppm_omp1_ad_acc.txt};
			\end{axis}
			
			\begin{axis}[
			ybar stacked, bar width=10pt, bar shift = 6pt,
			ymin=0, ymax=75,
			xmin=-.5, xmax=3.5,
			width=7cm,
			height=5cm,
			xtick={0,1,2,3},
			xticklabels={5e-5, 1e-4, 2e-4, 4e-4},
			y axis line style={opacity=0},				
			]				
	
				\pgfplotsinvokeforeach{0,1,2,3,4}{
	\node[anchor=south, white, scale=0.6] (A) at ($ (axis cs:#1,0.0) + 1*(6pt, 0)$) {\sc Opt};
	\node[anchor=south, white, scale=0.6] (A) at ($ (axis cs:#1,0.0) - 1*(6pt, 0)$) {\sc Ref};
}

			\addplot[color=black, fill=blue!90!green] table[x expr=\coordindex, y index=0, header=false]{img/pppm_new1_ad_acc.txt};
			\addplot[color=black, fill=blue!30!green] table[x expr=\coordindex, y index=1, header=false]{img/pppm_new1_ad_acc.txt};		
			\addplot[color=black, fill=blue!60!green] table[x expr=\coordindex, y index=2, header=false]{img/pppm_new1_ad_acc.txt};
			\addplot[color=black, fill=yellow!60!green] table[x expr=\coordindex, y index=3, header=false]{img/pppm_new1_ad_acc.txt};
			\end{axis}
\end{scope}
			\end{tikzpicture}				
		\vspace{-0.5cm}
		\caption{Profiles of SPC/E water test case running
                  single-threaded on one core of a KNL as PPPM relative error
                  varies.  Left bar: reference, right bar: optimized.}
                \label{fig_1xacc_compare}
		\end{figure}

		Fig. \ref{fig_1xacc_compare} compares the
		optimized implementation to the reference as relative error varies.  Speedups
		are between 2.1x and 2.77x for all cases, without an apparent pattern other
		than that {\sc ad} differentiation has gained more from the optimizations than {\sc ik} differentiation.  There is not a clear optimal relative error, since users will want to adjust this parameter depending on how important accuracy is in the long-range calculation for their specific problems.

		\subsection{OpenMP and MPI Parallelism}
		
		With the additional complication of parallelism, we do not attempt to
		determine optimal choices of input parameters for our test case, though users will go through this complex process for their individual problems, often settling on a suboptimal set of parameters~\cite{Diego16}.  Here we just show that our
		optimized version is much faster than the reference for a range of
		cutoffs on a full KNL node, for varying numbers of cores on up to two
		full nodes, and for varying numbers of OpenMP threads per
		rank.
		
		
		
		LAMMPS is intended to be scalable to very large numbers of cores, but
		this scalability is highly dependent on the details of the simulation.  As the number of MPI ranks increases, the
		runtime-optimal input parameters change.  Using just one set of input
		parameters might result in poor scalability (if the chosen set is optimal for small numbers of ranks) or good scalability (if the chosen set is optimal for a large number of ranks).  As the number of ranks grows, FFTs and other functions requiring communication become relatively more expensive.  This increases the runtime-optimal cutoff and can also make using a stencil size of 7 more efficient even for reference LAMMPS.  Parallelism provides yet more knobs for users to consider.  These include the number of MPI ranks per node and a number of OpenMP threads per rank.  The optimal choice is again problem-dependent, but generally LAMMPS should be run with around 1 core per rank and 1-2 threads per core.
		
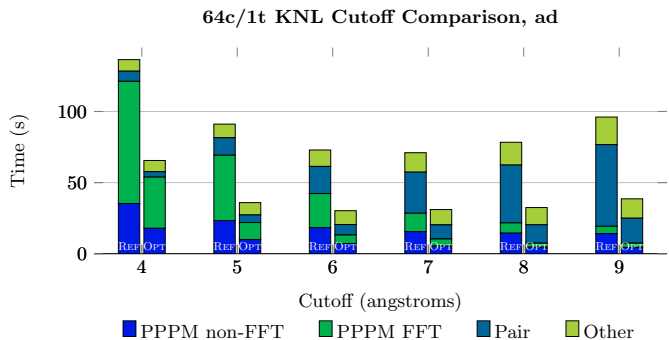
\begin{figure}
			\centering
			\begin{tikzpicture}[scale=0.8]
			\begin{axis}[
			ybar stacked, bar width=10pt, bar shift = -6pt,
			ymin=0, ymax=145,
			xmin=-.5, xmax=5.5,
			width=11cm,
			height=5cm,
			xtick={0,1,2,3,4, 5},
			xticklabels={4, 5, 6, 7, 8, 9},
			legend style={at={(0.5,-0.3)},anchor=north,/tikz/every even column/.append style={column sep=0.5cm}, anchor=north,draw=none, legend columns=4},
			ymajorgrids,
			ylabel={Time (s)},
			axis x line*=bottom,
			y axis line style={opacity=0},
			xlabel ={Cutoff (angstroms)},
			title={\textbf{64c/1t KNL Cutoff Comparison, ad}}
			]
			\addplot[color=black, fill=blue!90!green] table[x expr=\coordindex, y index=0, header=false]{img/pppm_omp64_ad_cut.txt};
			\addplot[color=black, fill=blue!30!green] table[x expr=\coordindex, y index=1, header=false]{img/pppm_omp64_ad_cut.txt};	
			\addplot[color=black, fill=blue!60!green] table[x expr=\coordindex, y index=2, header=false]{img/pppm_omp64_ad_cut.txt};
			\addplot[color=black, fill=yellow!60!green] table[x expr=\coordindex, y index=3, header=false]{img/pppm_omp64_ad_cut.txt};
			\legend{PPPM non-FFT, PPPM FFT, Pair, Other}				
			\end{axis}
			
			\begin{axis}[
			ybar stacked, bar width=10pt, bar shift = 6pt,
			ymin=0, ymax=145,
			xmin=-.5, xmax=5.5,
			width=11cm,
			height=5cm,
			xtick={0,1,2,3,4, 5},
			xticklabels={4, 5, 6, 7, 8, 9},
			y axis line style={opacity=0},				
			]	
				\pgfplotsinvokeforeach{0,1,2,3,4,5}{
	\node[anchor=south, white, scale=0.6] (A) at ($ (axis cs:#1,0.0) + 1*(6pt, 0)$) {\sc Opt};
	\node[anchor=south, white, scale=0.6] (A) at ($ (axis cs:#1,0.0) - 1*(6pt, 0)$) {\sc Ref};
}

			\addplot[color=black, fill=blue!90!green] table[x expr=\coordindex, y index=0, header=false]{img/pppm_new64_ad_cut.txt};
			\addplot[color=black, fill=blue!30!green] table[x expr=\coordindex, y index=1, header=false]{img/pppm_new64_ad_cut.txt};
			\addplot[color=black, fill=blue!60!green] table[x expr=\coordindex, y index=2, header=false]{img/pppm_new64_ad_cut.txt};
			\addplot[color=black, fill=yellow!60!green] table[x expr=\coordindex, y index=3, header=false]{img/pppm_new64_ad_cut.txt};
			\end{axis}
			\end{tikzpicture}				
			\vspace{-0.5cm}
			\caption{Profiles of SPC/E water test case scaled up
                          by 64x running single-threaded on a full KNL node
                          as cutoff varies, for {\sc ad} differentiation.  Left bar: reference, right bar: optimized.}
			\label{fig-64xcut_compare}
		\end{figure}

		Fig. \ref{fig-64xcut_compare} contains results for running a proportionally scaled-up benchmark on an entire KNL node with all 64 of its cores.  Now we present results for cutoffs from 4 to 9\AA\ instead of 3 to 7 \AA, since at 3\AA\ the FFTs for both versions take much longer.  For reference LAMMPS the runtime-optimal cutoff is now at 7\AA.  The optimized version is fastest at 6\AA, although 7\AA\ is only slightly slower.  
		This set of simulations features the same number of atoms per core as Fig. \ref{fig-1xadcut_compare}, but its efficiency is reduced by parallelism overhead.  For the single-core optimal cutoff of 5\AA, this scaled-up simulation takes 2.5 times as long per atom with our optimized code.  It still takes about twice as long even at the new optimal cutoff of 6\AA.  The reference version fares a little better, taking ``only" twice as long at 5\AA\ and 1.4 times as long at its new optimal cutoff of 7\AA.  
		If instead we compare the times required at the new runtime-optimal cutoffs to that required for the single-core optimal cutoff, the full node simulations take 1.8 and 2.1 times longer for the reference and optimized codes, respectively. Scalability aside, however, the same general patterns are apparent here as were seen earlier.  Total speedup is about 2.4x for optimal cutoffs, lower than for the single-core case due to the relative increase in the expense of communication-intensive functions.

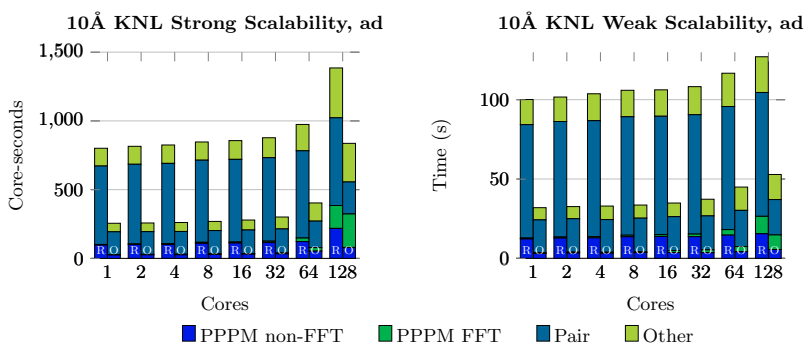
\begin{figure}
			\begin{tikzpicture}[scale=0.8]
			\begin{axis}[
			ybar stacked, bar width=6pt, bar shift = -3pt,
			ymin=0, ymax=1500,
			xmin=-.5, xmax=7.5,
			width=6cm,
			height=5cm,
			xtick={0,1,2,3,4, 5, 6, 7},
			xticklabels={1, 2, 4, 8, 16, 32, 64, 128},
			legend style={at={(1.3,-0.3)},anchor=north,/tikz/every even column/.append style={column sep=0.5cm}, anchor=north,draw=none, legend columns=4},
			ymajorgrids,
			ylabel={Core-seconds},
			axis x line*=bottom,
			y axis line style={opacity=0},
			xlabel ={Cores},
			title={\textbf{10\AA\ KNL Strong Scalability, ad}}
			]
			\addplot[color=black, fill=blue!90!green] table[x expr=\coordindex, y index=0, header=false]{img/pppm_omp_scalability.txt};
			\addplot[color=black, fill=blue!30!green] table[x expr=\coordindex, y index=1, header=false]{img/pppm_omp_scalability.txt};	
			\addplot[color=black, fill=blue!60!green] table[x expr=\coordindex, y index=2, header=false]{img/pppm_omp_scalability.txt};
			\addplot[color=black, fill=yellow!60!green] table[x expr=\coordindex, y index=3, header=false]{img/pppm_omp_scalability.txt};
			\legend{PPPM non-FFT, PPPM FFT, Pair, Other}				
			\end{axis}
			
			\begin{axis}[
			ybar stacked, bar width=6pt, bar shift = 3pt,
			ymin=0, ymax=1500,
			xmin=-.5, xmax=7.5,
			width=6cm,
			height=5cm,
			xtick={0,1,2,3,4, 5, 6, 7},
			xticklabels={1, 2, 4, 8, 16, 32, 64, 128},
			y axis line style={opacity=0},				
			]	
				\pgfplotsinvokeforeach{0,1,2,3,4,5,6,7}{
	\node[anchor=south, white, scale=0.6] (A) at ($ (axis cs:#1,0.0) + 1*(3pt, 0)$) {\sc O};
	\node[anchor=south, white, scale=0.6] (A) at ($ (axis cs:#1,0.0) - 1*(3pt, 0)$) {\sc R};
}

			\addplot[color=black, fill=blue!90!green] table[x expr=\coordindex, y index=0, header=false]{img/pppm_new_scalability.txt};
			\addplot[color=black, fill=blue!30!green] table[x expr=\coordindex, y index=1, header=false]{img/pppm_new_scalability.txt};
			\addplot[color=black, fill=blue!60!green] table[x expr=\coordindex, y index=2, header=false]{img/pppm_new_scalability.txt};
			\addplot[color=black, fill=yellow!60!green] table[x expr=\coordindex, y index=3, header=false]{img/pppm_new_scalability.txt};
			\end{axis}
			\begin{scope}[xshift=7cm]
			\begin{axis}[
			ybar stacked, bar width=6pt, bar shift = -3pt,
			ymin=0, ymax=130,
			xmin=-.5, xmax=7.5,
			width=6cm,
			height=5cm,
			xtick={0,1,2,3,4, 5, 6, 7},
			xticklabels={1, 2, 4, 8, 16, 32, 64, 128},
			legend style={at={(0.5,-0.2)},anchor=north,/tikz/every even column/.append style={column sep=0.5cm}, anchor=north,draw=none, legend columns=2},
			ymajorgrids,
			ylabel={Time (s)},
			axis x line*=bottom,
			y axis line style={opacity=0},
			xlabel ={Cores},
			title={\textbf{10\AA\ KNL Weak Scalability, ad}}
			]
			\addplot[color=black, fill=blue!90!green] table[x expr=\coordindex, y index=0, header=false]{img/pppm_omp_weakscalability.txt};
			\addplot[color=black, fill=blue!30!green] table[x expr=\coordindex, y index=1, header=false]{img/pppm_omp_weakscalability.txt};	
			\addplot[color=black, fill=blue!60!green] table[x expr=\coordindex, y index=2, header=false]{img/pppm_omp_weakscalability.txt};
			\addplot[color=black, fill=yellow!60!green] table[x expr=\coordindex, y index=3, header=false]{img/pppm_omp_weakscalability.txt};
			\end{axis}
			
			\begin{axis}[
			ybar stacked, bar width=6pt, bar shift = 3pt,
			ymin=0, ymax=130,
			xmin=-.5, xmax=7.5,
			width=6cm,
			height=5cm,
			xtick={0,1,2,3,4, 5, 6, 7},
			xticklabels={1, 2, 4, 8, 16, 32, 64, 128},
			y axis line style={opacity=0},				
			]	

				\pgfplotsinvokeforeach{0,1,2,3,4,5,6,7}{
	\node[anchor=south, white, scale=0.6] (A) at ($ (axis cs:#1,0.0) + 1*(3pt, 0)$) {\sc O};
	\node[anchor=south, white, scale=0.6] (A) at ($ (axis cs:#1,0.0) - 1*(3pt, 0)$) {\sc R};
}

			\addplot[color=black, fill=blue!90!green] table[x expr=\coordindex, y index=0, header=false]{img/pppm_new_weakscalability.txt};
			\addplot[color=black, fill=blue!30!green] table[x expr=\coordindex, y index=1, header=false]{img/pppm_new_weakscalability.txt};
			\addplot[color=black, fill=blue!60!green] table[x expr=\coordindex, y index=2, header=false]{img/pppm_new_weakscalability.txt};
			\addplot[color=black, fill=yellow!60!green] table[x expr=\coordindex, y index=3, header=false]{img/pppm_new_weakscalability.txt};
			\end{axis}
			\end{scope}
			\end{tikzpicture}
			\vspace{-0.5cm}		
			\caption{Strong and weak scalability comparisons up to 2 full KNL nodes.  Left bar: reference, right bar: optimized.}
			\label{fig-scalability}

		\end{figure}

		As it appears on the LAMMPS website, the SPC/E water benchmark we use here defaults to a cutoff of 9.8\AA.  This is of course far higher than the runtime-optimal cutoff on a single core---the simulation takes more than twice as long as at 5\AA\ for reference LAMMPS and about twice as long for our optimized version.  However, this exhibits much better scalability since runtime-optimal cutoffs are higher for higher core counts.  This is because less time is spent performing poorly-scaling FFTs while more time is spent computing short-range pair interactions.  Fig. \ref{fig-scalability}a shows core-seconds taken to simulate a fixed-size problem as the number of cores used increases.  There is one MPI rank per core and 1 thread per rank.  
		For the 10\AA\ case this scales well up to 32 cores, but for the full KNL node parallel efficiencies are 82\% for reference LAMMPS and 63\% for optimized LAMMPS.  Running on 128 cores across two full nodes is very inefficient; the optimized version actually runs faster on one node, in part due to using an unfortunate number of FFT grid points, although it remains faster than the reference.  Both versions see a comparable increase in core-seconds as communication costs rise, and this has a larger relative impact on the optimized version because it was faster to begin with.  These observations are consistent with benchmarks published on the LAMMPS website, which exhibit large losses in parallel efficiency after about 16 processors for a variety of systems when running fixed-size benchmarks.  

	More commonly, users will simulate large problems on large numbers of cores.  Fig. \ref{fig-scalability}b shows core-seconds per atom as problem size and core count both vary, such that there are 36k atoms per core in each simulation.  Parallel efficiencies on a full KNL node are now 85\% and 71\% for the reference and optimized versions, respectively, and 79\% and 60\% for two full nodes.  Again we see the optimized version scaling less well because the rise in communication costs with core count is roughly the same for both versions, but it remains 2-3x faster over the entire range.
		
\begin{figure}
\centering
			\begin{tikzpicture}[scale=0.8]
			\begin{axis}[
			ybar stacked, bar width=10pt, bar shift = -6pt,
			xmin=-.5, xmax=3.5,
			ymin=0, ymax=95,
			width=11cm,
			height=5cm,
			xtick={0,1,2,3},
			xticklabels={16/4, 32/2, 64/1, 64/2},
			legend style={at={(0.5,-0.3)},anchor=north,/tikz/every even column/.append style={column sep=0.5cm}, anchor=north,draw=none, legend columns=4},
			ymajorgrids,
			ylabel={Time (s)},
			axis x line*=bottom,
			y axis line style={opacity=0},
			xlabel ={MPI Ranks / Threads per Rank},
			title={\textbf{Full Node KNL Hybrid MPI OpenMP Comparison, ad}}
			]
			\addplot[color=black, fill=blue!90!green] table[x expr=\coordindex, y index=0, header=false]{img/pppm_omp_hybrid.txt};
			\addplot[color=black, fill=blue!30!green] table[x expr=\coordindex, y index=1, header=false]{img/pppm_omp_hybrid.txt};		
			\addplot[color=black, fill=blue!60!green] table[x expr=\coordindex, y index=2, header=false]{img/pppm_omp_hybrid.txt};
			\addplot[color=black, fill=yellow!60!green] table[x expr=\coordindex, y index=3, header=false]{img/pppm_omp_hybrid.txt};
			\legend{PPPM non-FFT, PPPM FFT, Pair, Other}				
			\end{axis}
			
			\begin{axis}[
			ybar stacked, bar width=10pt, bar shift = 6pt,
			ymin=0, ymax=95,
			xmin=-.5, xmax=3.5,
			width=11cm,
			height=5cm,
			xtick={0,1,2,3},
			xticklabels={16/4, 32/2, 64/1, 64/2},
			y axis line style={opacity=0},				
			]	
				\pgfplotsinvokeforeach{0,1,2,3,4}{
	\node[anchor=south, white, scale=0.6] (A) at ($ (axis cs:#1,0.0) + 1*(6pt, 0)$) {\sc Opt};
	\node[anchor=south, white, scale=0.6] (A) at ($ (axis cs:#1,0.0) - 1*(6pt, 0)$) {\sc Ref};
}

			\addplot[color=black, fill=blue!90!green] table[x expr=\coordindex, y index=0, header=false]{img/pppm_new_hybrid.txt};
			\addplot[color=black, fill=blue!30!green] table[x expr=\coordindex, y index=1, header=false]{img/pppm_new_hybrid.txt};		
			\addplot[color=black, fill=blue!60!green] table[x expr=\coordindex, y index=2, header=false]{img/pppm_new_hybrid.txt};
			\addplot[color=black, fill=yellow!60!green] table[x expr=\coordindex, y index=3, header=false]{img/pppm_new_hybrid.txt};
			\end{axis}
			\end{tikzpicture}				
\vspace{-0.5cm}
			\caption{Profiles of 64x-scale SPC/E water test case
                          running on a full KNL node, for {\sc ad}
                          differentiation, varying the number of MPI ranks and
                          OpenMP threads per rank.  The reference user-omp
                          implementation is on the left and our optimized
                          implementation is on the right. The reference cases
                          were run with a stencil size of 5 and the optimized
                          cases with a stencil size of 7.  Left bar: reference, right bar: optimized.}
			\label{fig-hybrid}
		\end{figure}
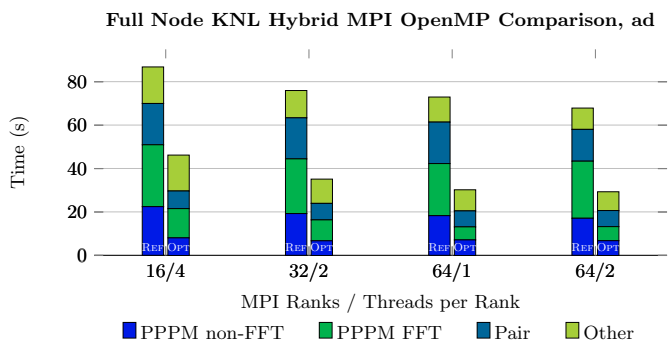

		Users can also make use of OpenMP parallelism, by either assigning multiple cores to each MPI rank or using multiple threads per physical core, or both.  Fig. \ref{fig-hybrid} shows profiles for the same 64x-scale water test case being simulated on a full KNL node, where the number of MPI ranks and OpenMP threads per rank is varied.  We use a cutoff of 6\AA, as this was close to the runtime-optimal cutoff for this case on the full node when using 64 MPI ranks and 1 thread per rank.  Best results are obtained when using one MPI rank per core, which is expected when not running on many nodes---the behavior on two full nodes is similar.  Slight performance gain is obtained by using two OpenMP threads per core, which helps a little when computing the short-range interactions.  The optimized version behaves similarly to the reference, and is at least twice as fast except when using too few MPI ranks.

		\section{Conclusion}

Efficient vectorization proved to be key to attaining significant speedups
over reference LAMMPS.  For the PPPM functions, we tested several approaches
and found memory access patterns to be particularly important.  However,
because the contiguous memory accesses were to be found in loops over stencil
points, the stencil size limited vectorization efficiency.  At the same time,
as other parts of the code were optimized, the FFTs became relatively more
expensive.  And from the beginning we were concerned with users having
difficulty choosing an optimal stencil size.  

All of these problems turn out
to have the same solution.  Targeting a higher default stencil size allowed whole rows of a larger stencil to be computed at once, enabling efficient vectorization.  Work shifted away from the FFTs and into newly-optimized functions when LAMMPS automatically adjusted the FFT grid to preserve accuracy.  And users who do not test a variety of stencil sizes are no longer missing out on potential performance, because $S=7$ is optimal for every case and can be made the  default.  The relatively more expensive FFTs also made another previously-hard choice much easier, as now {\sc ad} differentiation is significantly faster than {\sc ik} differentiation due to its requiring only half as many FFTs.  Although not discussed here, most of our optimizations are applicable to 256-bit vector registers and yield significant speedup on Xeon architectures, and similar speedup is also observed for different types of physical problems, such as an interfacial system where half of the domain is a vacuum.

LAMMPS is an extremely flexible program that allows and requires users to make
numerous choices when simulating their different physical problems, and our
optimized code is a significant improvement over reference LAMMPS, regardless
of a user's particular needs, for simulations which make use of the PPPM method
for electrostatics.  We achieve 2-3x speedup across a wide range of cutoff
radii, for different accuracy requirements of the long-range solver, for both
differentiation modes, and for different approaches to parallelization.

Many of these choices have a large impact on performance and even on simulation accuracy, often in ways that are not intuitive and not transparent to users as they try to work out how best to approach their problems.  Some, like the choice of stencil size, are sufficiently obscure that many users likely use the default value, some without even knowing that they even had a choice.  Other users will have gone to great lengths to set up their simulations in the best possible way, and will have made nearly-optimal choices for their specific problems.  Our optimizations are particularly helpful to the first group because several of the user-facing knobs now have clearly best settings for a range of problem sizes, and these settings can be clearly communicated without much qualification as to which cases they work for, or they can even be made the default selections.  Users with long experience and carefully-crafted input files will benefit from significant speedup for their existing set of inputs and can also expect that the optimal inputs for the new version are close to what they were already using.	
		
		


		\section*{Acknowledgments}

The authors gratefully acknowledge financial support from
the Deutsche Forschungsgemeinschaft (German Research Association) through
grant GSC 111, and from Intel Corporation via the Intel Parallel Computing Center initiative.

		
		
		\bibliographystyle{IEEEtran}
		\bibliography{IEEEabrv,mcdoniel_pppm_2017}

\end{document}